\documentclass[twocolumn,pra,amsmath,amssymb,showpacs]{revtex4}

\usepackage{bm}

\ifx\pdftexversion\undefined
  \usepackage[dvips]{graphicx}
\else
  \usepackage[pdftex]{graphicx}
\fi

\usepackage{color}

\def \be{\begin{equation}}
\def \ee{\end{equation}}
\def \bew{\begin{widetext}\begin{equation}}
\def \eew{\end{equation}\end{widetext}}
\def \bmlett{\begin{mathletters}}
\def \emlett{\end{mathletters}}

\def \pd{\phantom{\dagger}}

\def \sx{V_X}
\def \sy{V_Y}
\def \sxy{C}

\def \omegar{\omega_R}
\def \xrms{x_{\rm zpt}}
\def \chiR{\chi_R}
\def \chiM{\chi_M}
\def \bout{b_{\rm out}}
\def \bin{\hat{b}_{\rm in}}
\def \nbar{n_{\rm eq} }
\def \nBA{n_{\rm BA} }
\def \nBC{n_{\rm bad}}

\def \tnBA{\tilde{n}_{\rm BA} }

\def \tTeq{\tilde{T}_{\rm eq}}

\def \tkay{\tilde{k}}
\def \tlam{\tilde{\lambda}}
\def \tA{\tilde{A}}

\def \barx{\bar{X}}

\def \bary{\bar{Y}}

\def \ra{\rightarrow}

\def \hx{\hat{x}}

\def \hF{\hat{F}}

\def \hX{\hat{X}}
\def \hY{\hat{Y}}

\def \ha{\hat{a}}
\def \hc{\hat{c}}
\def \hH{\hat{H}}
\def \hd{\hat{d}}
\def \hf{\hat{f} }

\begin{document}




\title{Back-action evasion and squeezing of a mechanical resonator using a cavity detector}

\author{A. A. Clerk$^1$}
\author{F. Marquardt$^2$}
\author{K. Jacobs$^3$}
\affiliation{$^1$Department of Physics, McGill University, Montr\'{e}al,
 Qu\'{e}bec, Canada, H3A 2T8}
\affiliation{$^2$Department of Physics, Arnold-Sommerfeld-Center for Theoretical Physics, and Center for NanoScience, Ludwig-Maximilians-Universit\"at M\"unchen, Theresienstrasse 37, 80333 Munich, Germany}
\affiliation{$^3$Department of Physics, University of Massachussets at Boston, Boston, MA 02125, USA}

\date{Feb. 13, 2008}

\begin{abstract}

We study the quantum measurement of a cantilever using a parametrically-coupled
electromagnetic cavity which is driven at the two sidebands corresponding to the mechanical motion.  This scheme, originally due to Braginsky et al. [V. Braginsky, Y. I. Vorontsov, and K. P. Thorne, Science {\bf 209}, 547 (1980)],  allows a back-action free measurement of one quadrature of the cantilever's motion, and hence the possibility of generating a squeezed state.  We present a complete quantum theory of this system, and derive simple conditions on when the quantum limit on the added noise can be surpassed.  We also study the conditional dynamics of the measurement, and discuss how such a scheme (when coupled with feedback) can be used to generate and detect squeezed states of the oscillator.  Our results are relevant to experiments in optomechanics, and to experiments in quantum electromechanics employing stripline resonators coupled to mechanical resonators.

\end{abstract}


\maketitle

\section{Introduction} 

Considerable effort has been devoted recently to attempts at seeing quantum effects in micron to nanometre scale mechanical systems.  Experiments coupling such oscillators to mesoscopic electronic position detectors have seen evidence of quantum back-action and back-action cooling \cite{Naik06}, and have demonstrated continuous position detection at a level near the fundamental limit placed by quantum backaction \cite{Knobel03, LaHaye04, FlowersJacobs07}.  Complementary to this work, experiments using optomechanical systems (e.g.~a cantilever coupled to a optical cavity) have been able to cool micromechanical resonators by several orders of magnitude, using either passive \cite{Karrai04, Gigan06, Kippenberg06,Harris07} or active (i.e.~feedback-based) approaches \cite{Bouwmeester06} .  

Despite these recent successes, seeing truly quantum behaviour in a mechanical resonator remains a difficult challenge.  If one is only doing linear position detection, the quantum behaviour of an oscillator is almost perfectly masked.  Non-linear detector-oscillator couplings allow one to probe quantum behaviour such as energy quantization \cite{Santamore04,Jacobs07b,Martin07,Buks08,Jacobs08}; however, generating such couplings is generally not an easy task.  Quantum behaviour could also be revealed by coupling the resonator to a qubit \cite{Armour02, Wei06, Clerk07a}; this too is challenging, as it requires relatively large couplings and a highly phase coherent qubit.  Here, we consider an alternate route to seeing quantum behaviour in a mechanical oscillator, one that requires no qubit and only a linear coupling to position. As was first suggested by Braginsky and co-workers \cite{Braginsky80, Braginsky92}, by using an appropriately driven electromagnetic cavity which is parametrically coupled to a cantilever, one can make a measurement of just a single quadrature of the cantilever's motion.  As a result, quantum mechanical back-action need not place a limit on the measurement accuracy, as the back-action affects only the unmeasured quadrature.  One can then make (in principle) a perfect measurement of one quadrature of the oscillator's motion.  This is in itself useful, as it allows for the possibility of ultra-sensitive force detection \cite{Caves80b, BockoRMP96}.
Perhaps even more interesting, one expects that such a measurement can result in a quantum squeezed state of the oscillator, where the uncertainty of the measured quadrature drops below its zero point value. 

While the original proposal by Braginsky is quite old, there nonetheless does not exist a fully quantum theory of the noise and back-action of this scheme; moreover, there exists no treatment of the measurement-induced squeezing.  In this paper, we remedy this situation, and present a fully quantum theory of measurement in this system.  We calculate the full noise in the homodyned output signal from the cavity (an experimentally measurable quantity), and derive simple but precise conditions that are needed to beat the conventional quantum limit on the added noise of a position detector \cite{Caves80b,Braginsky92,Caves82,Clerk04c}.  Using a conditional measurement approach, we also discuss the conditions required to squeeze the mechanical resonator, and demonstrate how feedback may be used to unambiguously detect this squeezing.
Our results are especially timely, given the recent experimental successes in realizing cavity position detectors using both superconducting stripline resonators \cite{Regal08} as well as optical cavities \cite{Karrai04, Gigan06, Kippenberg06, Harris07, Kippenberg07}; our theory is applicable to both these classes of systems.    Note that Ruskov et al.~\cite{Ruskov05} recently analyzed a somewhat related scheme involving stroboscopic measurement of an oscillator with a quantum point contact.  Unlike that scheme, the system analyzed here should be much easier to implement, being directly related to existing experimental setups; our scheme also has the benefit of allowing significant squeezing without the need to generate extremely fast pulses.

The remainder of this paper is organized as follows.  In Sec.~\ref{sec:Main} we give a heuristic description of how one may realize back-action free single-quadrature detection, introduce the Braginsky two-sideband scheme, and give a synopsis of our main findings.  In Sec.~\ref{sec:Details} we provide the details of our calculations, and Sec.~\ref{sec:Conc} concludes.

\section{Model and Main Results}
\label{sec:Main}

\subsection{Basic idea behind single quadrature detection}
\label{subsec:BasicIdea}

Consider a high-Q mechanical oscillator having frequency $\omega_M$ and annihilation operator $\hc$.  We will use $X$ and $Y$ to denote the cosine and sine quadratures of the oscillator's motion.  Using Schr\"odinger operators $\hc$ and $\hc^{\dag}$, the operators associated with the quadratures are:
\begin{subequations}
\begin{eqnarray}
    \hat{X} & = & \frac{1}{\sqrt{2}} \left(
         \hat{c} e^{i \omega_M t} +  \hat{c}^\dag e^{-i \omega_M t}
    \right) \\
    \hat{Y} & = & -\frac{i}{\sqrt{2}} \left(
         \hat{c} e^{i \omega_M t} -  \hat{c}^\dag e^{-i \omega_M t}
    \right)
\end{eqnarray}
\label{eq:QuadOps}
\end{subequations}
The Heisenberg-picture position operator $\hx(t)$ is then given by the Heisenberg picture operators $\hX(t)$ and $\hY(t)$ via:
\begin{eqnarray}
	 \hx(t) \equiv
	 	\sqrt{2} \xrms \left(
	 	\hX(t) \cdot \cos \omega_M t + \hY(t) \sin \omega_M t 
		\right)
\end{eqnarray}
as expected.  Note that $\hX$ and $\hY$ are canonically conjugate:
\begin{eqnarray}
	\left[ \hX, \hY \right] = i
\end{eqnarray}
Also note that the definition of the quadrature operators relies on having an external clock in the system which defines the zero of time.  

In general, $\hX(t)$ and $ \hY(t) $ will vary slowly in time (in comparison to $\omega_M$) due to the external forces acting on the oscillator.  Our goal will be to make a weak, continuous measurement of {\it only} $\hX(t)$, using the usual kind of setup where the position of the oscillator is linearly coupled to a detector.  We will use a detector - oscillator coupling Hamiltonian of the form:
\begin{eqnarray}
	\hH_{\rm int} = -A \hx \hF
	\label{eq:GenericHInt}
\end{eqnarray}
where $\hF$ is some detector operator.  It represents the force exerted by the detector on the oscillator; in the cavity-position detector we will consider, $\hF$ will be the number of photons in an electromagnetic cavity.  Ideally, if we only measure $\hX$, the back-action of the measurement will only affect the unmeasured quadrature $\hY$, and will not affect the evolution of $\hX$ at later times.  The hope thus exists of being able to make a back-action free measurement, one which is not subject to the usual standard quantum limit \cite{Caves80b,Caves82,Clerk04c}.

As is discussed extensively in Refs.~\onlinecite{Caves80b} and \onlinecite{BockoRMP96}, single quadrature detection with the interaction Hamiltonian in Eq.~(\ref{eq:GenericHInt}) can be accomplished by simply modulating the coupling strength $A$ at the oscillator frequency.  Setting $A=A(t) =  2 (\tilde{A} / \xrms) \cos( \omega_M t) $, $\hH_{\rm int}$ becomes:
\begin{eqnarray}
	\hH_{\rm int} &= & - \sqrt{2} \tilde{A}  \hF \cdot 
		\nonumber \\
		&& \left[
		\hX \left(1 + \cos(2 \omega_M t) \right) + \hY \sin(2 \omega_M t) \right]
		\label{eq:RotatingHInt}
\end{eqnarray}
In a time-averaged sense, we see the detector is only coupled to the $X$ quadrature;
we thus might expect that the (time-averaged) output of the detector will tell us only about $X$.  In principle, this in itself does not imply a lack of back-action:  via the coupling to $\hY$, noise in $\hF$ could affect the dynamics of $\hX$.  To prevent this, we need the further requirement that {\it the detector force has no frequency components near $\pm 2 \omega_M$}.  In this case, the effective back-action force $\hF \sin(2 \omega_M t)$ will have no Fourier weight in the narrow-bandwidth around zero frequency to which $\hX$ is sensitive, and it will not affect $\hX$.  Note that Ruskov et al. \cite{Ruskov05} recently considered a linear position detection scheme where the effective coupling constant is harmonically modulated; however, their scheme does not satisfy the second requirement above of having a narrow-band back-action force.

\subsection{Model}

\begin{figure}
\begin{center}
\includegraphics[width=1.0 \columnwidth]{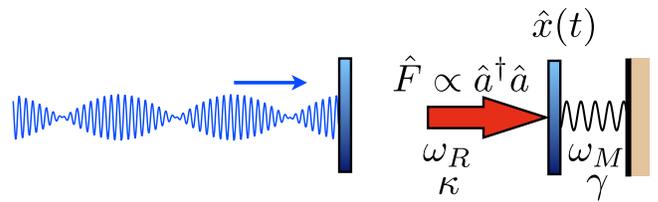}
\caption{Schematic picture of the setup studied in the text.
A cavity is driven by an input beam that is
amplitude-modulated at the mechanical frequency $\omega_M$ of
the movable end-mirror. The radiation pressure force,
as well as the cavity and mechanical frequencies and
decay rates are indicated.} 
\label{fig:SetupFigure}
\end{center}
\end{figure}
We now consider a specific and experimentally-realizable system which can realize the above ideas; this system was first proposed by Braginsky \cite{Braginsky80,Braginsky92}.  As shown schematically in Fig.~\ref{fig:SetupFigure}, the setup consists of a  high-Q mechanical oscillator  which is parametrically coupled with strength $A$ to a driven electromagnetic cavity:
\begin{eqnarray}
	\hat{H} & = & 
		\hbar(\omega_R - A \hx) 
			\left[ 
				\ha^{\dag} a  - \langle  \ha^{\dag} a \rangle 
			\right] \nonumber \\
		& & +	\hH_M + \hH_{\rm drive} + \hH_{\kappa} + \hH_{\gamma} 
\end{eqnarray}
where $\omega_R$ is the cavity resonance frequency, $\hH_M = \hbar \omega_M \hc^{\dag} \hc$ is the mechanical oscillator Hamiltonian, $\hH_{\rm drive}$ describes the cavity drive, $\hH_{\kappa}$ describes the cavity damping, and $\hH_{\gamma}$ is the mechanical damping.  Note this same system was recently shown (in a similar parameter regime to what we will require) to allow back-action cooling to the ground state \cite{MarquardtClerk07}.

Assuming a one-sided cavity, standard input-output theory \cite{WallsMilburn, Gardiner00} yields the Heisenberg equation of motion:
\be
	\dot{ \ha } = \left(-i \omega_R - \frac{\kappa}{2} \right) \ha
	 -\sqrt{\kappa}\, \bin(t).
\label{eq:cavityEOM3}
\ee
where $\kappa$ is the cavity damping, and $\bin$ describes both the drive applied to the cavity as well as the noise (quantum and thermal) entering the vacuum port.

To implement back-action evasion in the cavity system, we will
consider the case where $\omegar \gg \omega_M$, and
take the resolved-sideband or  ``good cavity" limit, where $\omega_M \gg \kappa$. 
We will also take an amplitude-modulated cavity drive of the form:
\begin{eqnarray}
    \langle \bin(t) \rangle \equiv \bar{b}_{\rm in}(t) & = &
        \frac{\bar{b}_{LO}}{2} \sin(\omega_M t) e^{-i \omegar t}
        \label{eq:aeq}
\end{eqnarray}
The same resolved sideband limit is required to achieve ground state cooling \cite{MarquardtClerk07}; all that is different from the setup here is the nature of the drive.  Here one drives the cavity equally at both sidebands associated with the oscillator motion, while in the cooling case, one only drives the red-detuned sideband.

To proceed, we may write the cavity
annihilation operator $\hat{a}$ as the sum of a classical piece
$\bar{a}(t)$ and a quantum piece $\hat{d}$:
\be
\hat{a}(t) = \bar a(t) + \hat{d}(t)
\label{eq:DisplacementTrans}
\ee  
$\bar{a}(t)$ is determined solely by the the response of the cavity to the (classical) external drive $\bar{b}_{\rm in}(t)$.  
In the long-time limit, Eq.~(\ref{eq:aeq})
yields:
\begin{eqnarray}
    \bar{a}(t) & = &
            \bar{a}_{\rm max} 
            \cos \left( \omega_M t
             + \delta \right) e^{-i \omegar t} 
\end{eqnarray}
with
\begin{subequations}
\begin{eqnarray}
	\bar{a}_{\rm max} & = & 
		\bar{b}_{LO} 
			\sqrt{ 
				\frac{ \kappa }{ 4 \omega_M^2 + \kappa^2 }
			} 	\\
	\delta & = & \arctan \left( \kappa / \omega_M \right)
\end{eqnarray}
\end{subequations}
The phase $\delta$ plays no role except to set the definitions of the two quadratures $X$ and $Y$; thus, without loss of generality, we will set it to zero.  We will also be interested in a drive large enough that $\bar{a}_{\rm max} \gg 1$.

In contrast to $\bar{a}$, $\hat{d}$, the quantum part of the cavity annihilation operator, is influenced by both the mechanical oscillator and quantum noise associated with the cavity dissipation.  Making use of the solution for $\bar{a}$ and the conditions $\omegar \gg \omega_M \gg
\kappa$, and keeping only terms which are at least order $\bar{a}$, the 
term in the total system Hamiltonian coupling the oscillator to the cavity takes an  analogous form to  Eq.~(\ref{eq:RotatingHInt}) with:
\begin{subequations}
\begin{eqnarray}
	\tilde{A} & = & \frac{1}{2} \left(A \xrms\right) \bar{a}_{\rm max}  \\
	\hat{F} & = & e^{i \omegar t} \hd + e^{-i \omegar t} \hd^{\dag}
	\label{eq:CavityForceOp}
\end{eqnarray}
\end{subequations}
Thus, the chosen cavity drive gives us the required harmonically-modulated coupling constant: in a time-averaged sense, the cavity is only coupled to the $X$ oscillator quadrature.  Further, the second condition outlined in Sec.~\ref{subsec:BasicIdea} is also satisfied:  because $\kappa \ll \omega_M$,  $\hF$ has no appreciable noise power at frequencies near $\pm 2 \omega_M$.  As such, we expect no back-action heating of the $\hX$ quadrature in the resolved-sideband limit $\kappa / \omega_M \ra 0$.  We will of course consider the effect of a non-zero but small $\kappa / \omega_M$ in what follows.
Note that electromechanical realizations of the system presented here, which use superconducting stripline resonators, can easily achieve the resolved sideband limit
$\omega_M \gg \kappa$ (e.g. the recent experiment of Ref.~\onlinecite{Regal08} achieved $\omega_M / \kappa \approx 5$).  The resolved sideband limit has also recently been reached for an optical mode propagating in a microtoroidal cavity that supports mechanical vibrations, with $\omega_M / \kappa \approx 20$ \cite{Kippenberg07}.

\subsection{Back-action}

Working in an interaction picture, one can easily derive Heisenberg equations of motion for the system, and solve these in the Fourier domain (c.f. Eqs.~(\ref{eq:XYSols})).  As expected, one finds that in the ideal good cavity limit ($\kappa / \omega_M \ra 0$), the measured $X$ quadrature is completely unaffected by the coupling to the cavity, while the unmeasured $Y$ quadrature experiences an extra back-action force due to the cavity.  For finite $\kappa / \omega_M$, there is some small additional back-action heating of the $X$ quadrature.  The noise spectral density of the quadrature fluctuations are given by:
\begin{subequations}
\begin{eqnarray}
	S_{X}(\omega) & \equiv &\frac{1}{2}
		\int_{-\infty}^{\infty} dt e^{i \omega t} 
			\left \langle \{ \hX(t), \hX(0) \} \right \rangle \nonumber \\
		& = &
		\frac{\gamma/2}{ \omega^2 + (\gamma/2)^2 } 
			\left[ 1 + 2 \left( \nbar + \nBC \right) \right]
			\label{eq:SXX} \\
	S_{Y}(\omega) & = &
		\frac{1}{2}
		\int_{-\infty}^{\infty} dt e^{i \omega t} 
			\left \langle \{ \hY(t), \hY(0) \} \right \rangle \nonumber \\
		& = &
		 \frac{\gamma / 2}
		 { \omega^2 + (\gamma/2)^2 } \left[
		 	1 + 2 \left( \nbar + \nBA + \nBC \right) \right] 
			\nonumber \\
		 \label{eq:SYY}
\end{eqnarray}
	\label{eq:NoiseSpectra}
\end{subequations}
where
\begin{equation}
	\nbar = \left( \exp\left[ \frac{\hbar \omega_M}{k_B T} \right] -1 \right)^{-1}
	\label{eq:nbar}
\end{equation}
is the number of thermal quanta in the oscillator.  $\nBA$ parameterizes the back-action heating of the $Y$ quadrature as an effective increase in $\nbar$; 
in the relevant limit $\gamma \ll \kappa$ one has:
\begin{eqnarray}
	n_{\rm BA} &= & \frac{ 8 \tilde{A}^2}{\kappa \gamma} 
	 =  
	 \frac{ 2\left(A \xrms \right)^2 }{\kappa \gamma}\left(\bar{a}_{\rm max} \right)^2
	\label{eq:nBADefn}
\end{eqnarray}
We have assumed here that the there is no thermal noise in the cavity drive: it is shot noise-limited.

Finally, $\nBC$ parameterizes the spurious back-action heating of $X$ which occurs when one deviates from the good-cavity limit; to leading order in $\kappa / \omega_M $, it is simply given by:
\begin{eqnarray}
	\nBC = \frac{\nBA}{32} \left( \frac{\kappa}{\omega_M} \right)^2
\end{eqnarray}

Note that there is no back-action damping of either quadrature (see discussion following Eqs.~(\ref{eq:XYSols})).

\subsection{Output Spectrum and Beating the SQL}

We assume that a homodyne measurement is made of the light leaving the cavity.
Using the solution to the Heisenberg equations of motion (c.f. Eqs.~(\ref{eq:XYSols})) and standard input-output theory, one can easily find the noise spectral density of the homodyne current $I(t)$.
The information about $\hX(t)$ will be contained in a bandwidth $\sim \gamma \ll \kappa$ around zero frequency.  Thus, focusing on frequencies $\omega \ll \kappa$, we have simply:
\begin{eqnarray}
	S_{I}(\omega) = G^2 \left[ S_{X}(\omega) +
		\frac{\kappa}{ 32 \tilde{A}^2}  S_{0}  \right]
		\label{eq:HomoSpectrum}
\end{eqnarray}
Here, $G$ is a gain coefficient proportional to the homodyne local oscillator amplitude, and $S_{0}$ represents added noise in the measurement coming from both the cavity drive and in the homodyne detection.  If both are shot noise limited, we simply have $S_{0} = 1$.  We can refer this noise back to the oscillator by simply dividing out the factor $G^2$:  the result is the {\it measured} $X$ quadrature fluctuations:
\begin{eqnarray}
		S_{X,{\rm meas} }(\omega) \equiv \frac{S_I(\omega)}{G^2}	
	=  S_{X}(\omega) +
		\frac{\kappa}{ 32 \tilde{A}^2}  S_{0}  
		\label{eq:SXXmeas}
\end{eqnarray}
Now, note that in the good cavity limit the spurious heating of $X$ described by $\nBC$ vanishes. Thus, in this limit, the added noise term (second term in Eq.~(\ref{eq:SXXmeas})) can be made arbitrarily small by increasing the intensity of the cavity drive beam (and hence $\tilde{A}$), without any resulting back-action heating of the measured $X$ quadrature.  Thus, in the good-cavity limit, there is no back-action imposed limit on how small we can make the added noise of the measurement (referred back to the oscillator).  In contrast, for small but non-zero $\kappa / \omega_M$, one needs to worry about the small residual back-action described by $\nBC$; one can still nonetheless beat the standard quantum limit in this case, as we now show.

To compare against the standard quantum limit, consider $S_{X,{\rm meas}}(0)$:
\begin{eqnarray}
	S_{X,{\rm meas} }(0) & = &  
		\frac{2}{\gamma}\left(1 + 2 \nbar  + 2 \nBC \right) +
			\frac{\kappa}{ 32 \tilde{A}^2}  S_{0} \nonumber \\
	& \equiv &
		\frac{2}{\gamma}\left(1 + 2 \nbar + 2 n_{\rm add} \right)
		\label{eq:SXUCzero}
\end{eqnarray}
In the last line, we have represented both the residual back-action $\nBC$ and the added noise of the measurement as an effective increase in the number of oscillator quanta by an amount $n_{\rm add}$.  The standard quantum limit (which applies when both quadratures are measured) yields the condition $n_{\rm add} \geq 1/2$ \cite{Caves80b, Caves82, Clerk04c}.  Here, we find:
\begin{eqnarray}
	n_{\rm add} = 
		\nBC + \frac{\kappa \gamma}{ 128 \tilde{A}^2}  S_{0}
		= \frac{\nBA}{32}\left( \frac{\kappa}{\omega_M}\right)^2 + \frac{1}{16 \nBA} S_{0} \nonumber \\
	\label{eq:nadd}
\end{eqnarray} 
Thus, if we are in the ideal good-cavity limit ($\kappa/ \omega_M \ra 0$) and shot-noise limited, beating the standard quantum limit on $n_{\rm add}$ requires a coupling strong enough that $\nBA \geq 1/8$:  the $Y$ quadrature fluctuations must be heated up by at least an eighth of an oscillator quantum. 

\begin{figure}
\begin{center}
\includegraphics[width=1.0 \columnwidth]{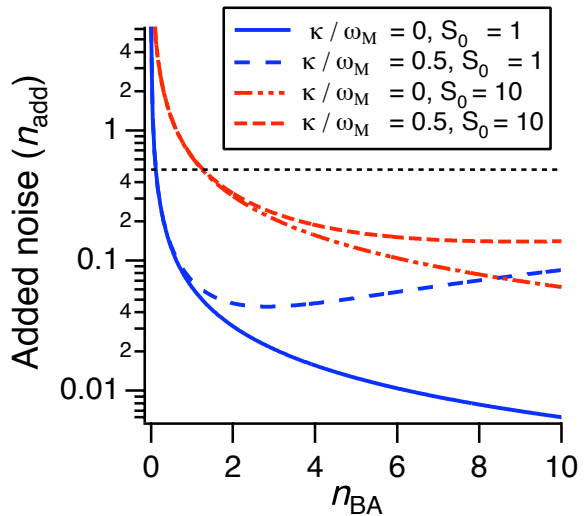}
\caption{Plot of added noise in the single quadrature measurement (measured as a number of quanta, $n_{\rm add}$, c.f.~Eq.~(\ref{eq:nadd})) versus the strength of the measurement (measured in terms of the back-action heating of the $Y$ quadrature, $\nBA$, c.f.~Eq.~\ref{eq:nBADefn}).  Different curves correspond to different values of $\kappa / \omega_M$ and $S_0$, the noise associated with the homodyne measurement; $S_0=1$ corresponds to a shot-noise limited measurement.  The standard quantum limit of $n_{\rm add} = 0.5$ is shown as a horizontal dashed line.  } 
\label{fig:AddedNoisePlot}
\end{center}
\end{figure}

In the more general case where $\kappa / \omega_M$ is finite, one cannot increase the coupling indefinitely, as there is back-action on $\hX$.  One finds that for an optimized coupling of:
\begin{eqnarray}
	\tnBA = \frac{ \omega_M}{\kappa} \sqrt{2 S_0}
\end{eqnarray}
the minimum added noise at resonance is given by:
\begin{eqnarray}
	n_{\rm add} \Big|_{min} =  \frac{\kappa}{8 \omega_M} 
		\sqrt{ \frac{S_0}{2}}
\end{eqnarray}
Thus, even for moderately small $\kappa / \omega_M$, one can make $n_{\rm add}$ smaller than the standard quantum limit value (see Fig.~\ref{fig:AddedNoisePlot}).

\subsection{Conditional Squeezing}

Given that the double-sideband scheme described here can allow for a near perfect measurement of the oscillator $X$ quadrature, one would expect it could lead to a squeezed oscillator state, where the uncertainty in $\hX$ drops below the zero point value of $1/2$.  However, Eq.~(\ref{eq:SXX}) indicates that in the good cavity limit, the fluctuations of $\hX$ are {\it completely} unaffected by the coupling to the cavity detector.  To resolve this seeming contradiction, one must consider the {\it conditional} aspects of the measurement: what is the state of the resonator in a particular run of the experiment?  In any given run of the experiment, the oscillator will indeed be squeezed.  However, the mean value of $\hX$ will have some non-zero value which is correlated with the noise in the output signal.  Once one averages over many realizations of the experiment, this random motion of $\langle \hX \rangle$ appears as extra noise, and masks the squeezing, resulting in the result of Eq.~(\ref{eq:SXX}).  We make these statements precise in what follows.

A rigorous description of the conditional evolution of the oscillator in the setup considered here can be developed in analogy to Ref.~\cite{DJ}, which considered ordinary linear position detection using a cavity.  For simplicity, we focus on the good cavity limit, where $\kappa / \omega_M \ra 0$.  We first define the parameter $\tkay$, a measure of the rate at which the measurement extracts information, as:
\begin{eqnarray}
	\tkay = \eta \frac{ 32  \tilde{A}^2}{\kappa}
		= \eta  \left( 4 \gamma \nBA \right) 
		\label{eq:tkay}
\end{eqnarray}
where $\nBA$ represents as before the back-action heating of the $Y$ quadrature, and $\eta = \frac{1}{S_0} \leq 1$ represents the efficiency of the homodyne detection ($\eta = 1$ corresponds to being quantum limited).  One has $\tkay = 1 / \tau_{\rm meas}$, where $\tau_{\rm meas}$ is the minimum time required to resolve a difference in $\langle X\rangle$ equal to the zero point rms value from the output of the detector; as we are interested in weak measurements, we expect $\tkay / \omega_M \ll 1$. Note that $\tkay = 8\eta k$, where $k$ is the usual definition of the strength of the measurement~\cite{JacobsSteck06}. 
The scaled homodyne output signal may then be written~\cite{JacobsSteck06}:
\begin{eqnarray}
	I(t) = \sqrt{ \tkay} \langle \hX(t) \rangle + \xi(t)
\end{eqnarray}
where $\xi(t)$ is white Gaussian noise.  In a given run of the experiment, $\xi(t)$ will be correlated with the state of the oscillator at times {\em later} than $t$.

In exact analogy to Ref.~\onlinecite{DJ}, a simple description of the conditional density matrix is possible in the limit where $\kappa \gg \tilde{A}$.  In this limit, the oscillator density matrix is Gaussian, being fully determined by its means $\bar{X} = \langle \hX \rangle, \bar{Y} = \langle \hY \rangle$ and its second moments $V_X = \langle \langle \hX^2 \rangle \rangle, V_Y = \langle \langle \hY^2 \rangle \rangle$ and $C = \langle \langle \{\hX, \hY\}/2 \rangle \rangle$.  In the interaction picture (i.e.~rotating frame at the oscillator frequency), the equations for the means (the estimates) are 
\begin{subequations}
\begin{eqnarray}
      \dot{ \barx }  & = & - \frac{\gamma}{2} \barx + \sqrt{\tkay} V_X \xi  \label{eq:est1}\\
	\dot{ \bary }  & = & - \frac{\gamma}{2} \bary + \sqrt{\tkay} C \xi
\end{eqnarray}
and for the covariances are 
\begin{eqnarray}
	\dot{ V}_X  & = & - \tkay V_X^2 - \gamma (V_X - \tTeq)   \label{eq:SxDot}\\
	 \dot{V}_Y & = & - \tkay C^2  + \tkay/(4\eta) - \gamma( V_Y - \tTeq) \\  
	\dot{C}\; & = & - \gamma C  - \tkay V_X C   \label{eq:est5}
\end{eqnarray}
\end{subequations}
where:
\begin{eqnarray}
	\tTeq = \frac{1}{2} + \nbar
\end{eqnarray}
We stress that these equations are almost identical to the standard equations for conditional linear position detection \cite{DJ, Ruskov05}, with the important exception that terms corresponding to the bare oscillator Hamiltonian are missing.  In a sense, the scheme presented here effectively transforms away the oscillator Hamiltonian.  

\begin{figure}
\begin{center}
\includegraphics[width=1.0 \columnwidth]{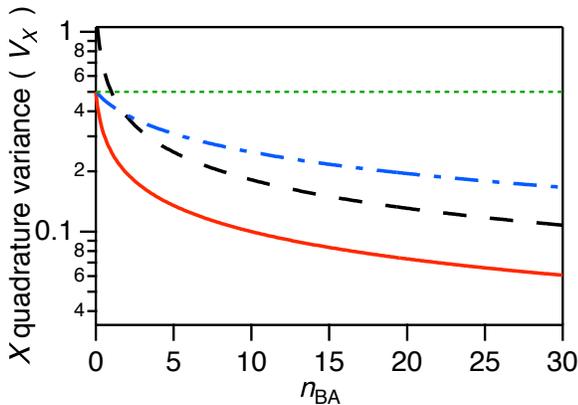}
\caption{Plot of the {\it conditional} X-quadrature variance $V_X$ (c.f.~Eq.~(\ref{eq:VXsoln})) as a function of the measurement strength (parametrized in terms of the back-action heating of the $Y$ quadrature, $\nBA$, c.f.~Eq.~(\ref{eq:nBADefn})); one clearly sees that the $X$-quadrature can be squeezed.  Different curves correspond to different values of the bath temperature (parameterized by $\nbar$, c.f.~Eq.~(\ref{eq:nbar})) and measurement efficiency $\eta$.   The solid red curve corresponds to $\nbar = 0$ and $\eta = 1$; the dashed black curve to $\nbar=1$ and $\eta = 1$; the dashed-dot blue curve to $\nbar=0$, $\eta = 0.1$.  The horizontal dashed line corresponds to the ground state value of the variance, $V_X = 0.5$.}
\label{fig:VXPlot}
\end{center}
\end{figure}

To find the amount of squeezing in a particular run of the experiment, we simply find the stationary variances for the oscillator's Gaussian state.  We have:
\begin{subequations}
\begin{eqnarray}
	V_Y & = & \frac{1}{2} +  \nbar + \nBA  \\
	V_X & = & \frac{\sqrt{2 \left(1 + 2 \nbar \right) \left(\eta \nBA \right) + 1/4}-1/2}
		{4 \left(\eta \nBA \right)}  \label{eq:VXsoln} \\
	C & = & 0
\end{eqnarray}
\end{subequations}
Note first that the result for $\sy$ is in complete agreement with the unconditional result of Eq.~(\ref{eq:SYY}): the measurement back-action heats the $Y$ quadrature by an amount corresponding to $\nBA$ quanta.  In contrast, we find that unlike the unconditional result of Eq.~(\ref{eq:SXX}), the measurement causes $\sx$ to decrease below its zero-coupling value:  it is a monotonically decreasing function of $\nBA$ (see Fig.~\ref{fig:VXPlot}).  This is the expected measurement-induced squeezing.  Of particular interest is the minimum coupling strength needed to reduce $\sx$ to its zero-point value:
\begin{eqnarray}
	\nBA = \frac{\nbar}{\eta}
\end{eqnarray}
In other words, lowering the $X$ quadrature uncertainty from a thermal value of $(1/2 +  \nbar)$ to the ground state value of $(1/2)$ requires that we {\it at least} increase the $Y$-quadrature uncertainty by the same amount.  This minimum amount is only achieved for a quantum-limited detector $\eta = 1$.  

The equation describing the fluctuations of the mean quadrature amplitudes $\bar{X},\bar{Y}$ can also be easily solved.  Assuming that $\barx = \bary =0$ at the initial time, one always has $\langle \barx(t) \rangle = \langle \bary(t) \rangle = 0$, where the average here is over many runs of the experiment.  In the stationary state (i.e.~once the variances $\sx, \sy$ and $\sxy$ have attained their stationary value), one finds $\bary(t) = 0$ with no fluctuations.  $\barx(t)$ continues to fluctuate, with an autocorrelation function:
\begin{eqnarray}
	\langle \barx(t) \barx(0) \rangle & = & 
		\left( \frac{ \tkay }{ \gamma} \right) V_X^2 e^{-\gamma |t| / 2} 
		\label{eq:XBarFlucs}
\end{eqnarray}
Again the average here is over many runs of the experiment.

We may now combine the results of Eq.~(\ref{eq:XBarFlucs}), and Eq.~(\ref{eq:SxDot}) to find $V_{X,{\rm tot}}$, the total (unconditional) $X$ variance.  One finds the simple result (valid in the stationary regime):
\begin{eqnarray}
	V_{X,{\rm tot}} & \equiv & 
		V_X + \langle \barx^2 \rangle 
			\nonumber \\
		&= & 
		V_X + \frac{\tkay}{\gamma} V_X^2 = 
		\frac{1}{2} + \nbar
	\label{eq:VXtot}
\end{eqnarray} 
This shows that, as expected, averaging the results of the conditional theory over many measurement runs reproduces the result of the unconditional theory (i.e.~the fluctuations of the measured $X$ quadrature are completely unaffected by the measurement). 

\subsection{Feedback for true squeezing}

In the previous section we saw how the state of the resonator, once conditioned 
on the measurement record, is squeezed. We can use feedback control to turn this conditional squeezing into ``real" squeezing of the resonator, where the full, unconditional oscillator variance $V_{X,{\rm tot}}$ (c.f Eq.~(\ref{eq:VXtot})) drops below the zero-point value.  This is accomplished by applying a time-dependent force to the resonator which is proportional to $\barx(t)$, the measured value of the $X$ quadrature.  Such a force can be used to suppress the fluctuations in the mean value of $X$, and in the limit of strong feedback, can remove them completely.
The only fluctuations that remain are quantified by the conditional variances, which are squeezed. Note that a similar approach was considered in Ref.~\cite{Ruskov05}.  

More precisely, if one makes the measurement at rate $\tkay$ described above (c.f. Eq.~(\ref{eq:tkay})), and applies the feedback force $F(t) =  \alpha \gamma \barx \sin \omega_M t$ in the laboratory frame, the result is an effective damping of the $X$ quadrature at a rate $\alpha \gamma /2$. Calculating the fluctuations of the $X$ quadrature under this feedback (the details of which are given in the next section), we find that the total unconditioned $X$ quadrature variance reaches a stationary state:
\begin{equation}
   V_{X,\rm tot}^{\mbox{\scriptsize fb}} = \frac{(1 + \alpha  + \frac{\tkay}{\gamma} V_X)V_X}{1 + \alpha}
   \label{eq:VXtotfb}   
\end{equation}
Here, $V_X$ is the conditioned variance given in Eq.~(\ref{eq:VXsoln}).  Note that when $\alpha \ra 0$, we again get the result of Eq.~(\ref{eq:VXtot}): the unconditioned $X$ quadrature variance is not affected by the measurement.  In contrast, in the limit of large $\alpha$, one has $ V_{X,\rm tot}^{\mbox{\scriptsize fb}}  \ra V_X$.
Thus, as claimed above, in the limit of strong feedback, the total fluctuations of the $X$ quadrature are reduced to the conditional variance; it may thus be squeezed.

It is also important to ask how this squeezing will manifest itself in the measurement signal. Calculating $S^{\mbox{\scriptsize fb}}_{I}(\omega)$, the spectrum of the homodyne current in the presence of feedback, and referring it back to the $X$ quadrature, we find:
 \begin{equation}
	 S^{\mbox{\scriptsize fb}}_{X,{\rm meas}} (\omega) \equiv
	 \frac{S^{\mbox{\scriptsize fb}}_{I}(\omega)}  {\tkay} =
	\frac{1}{\tkay} + \frac{(\nbar + 1/2) + \alpha V_X}{\omega^2 + 
		\left[ \frac{\gamma}{2}(1+ \alpha) \right]^2 }
   \label{eq:SI1fb}
\end{equation}
In the absence of feedback ($\alpha \ra 0$), the output spectrum consists of the white added noise of the measurement ($1/\tkay$) plus the measurement-independent $X$-quadrature fluctuation spectrum $S_X(\omega)$; this is in complete agreement with the unconditional theory (c.f.~Eq.~(\ref{eq:SXUCzero})).  When feedback is turned on, the second Lorentzian term in the spectrum is modified; this corresponds to the added damping and noise caused by the feedback.  In the limit of strong feedback, where $\alpha \ra \infty$, we find somewhat surprisingly that all signatures of the oscillator disappear: $S^{\mbox{\scriptsize fb}}_{X,{\rm meas}} (\omega) \ra 1/\tkay$.  Thus, in the limit of strong feedback, while the fluctuations in the $X$ quadrature are $V_X$, they do not appear at all in the output signal! To understand why this is, note that in driving the resonator with a force proportional $\barx$, we are driving it with a signal that is correlated with the noise in the output signal.  Thus, feedback leads to new {\em correlations} between the fluctuations of $X$ and the output noise.  In the limit of strong feedback, one finds that the fluctuations of $X$ have a variance $V_X$, but are perfectly 
{\em negatively} correlated with the output noise.  The result is that the output noise is completely independent of $V_X$.

The above effect, in which the fluctuations of $X$ vanish in the output signal, may be regarded as an example of {\em noise squashing} \cite{Buchler99, Wiseman99, Rugar07}. This is when one uses feedback specifically to reduce the fluctuations in the output signal, rather than to reduce the fluctuations of the system being measured.  This is possible only because the feedback uses the output signal, and thus correlates the system's fluctuations with the output fluctuations.  Strictly speaking, squashing refers only to the output signal that is part of the feedback loop (the so-called ``in-loop signal"), not to the actual system being measured.

While the existence of squashing in no way invalidates the real squeezing produced by the feedback, it does make it more difficult to observe this squeezing at the detector output. Further, any experimental results may be subject to the accusation that the feedback protocol may have been incorrectly designed to produce only squashing, with the result that squeezing of the resonator could not be inferred from the spectrum of the output. 

A solution to this problem is to make a second measurement of the mechanical resonator's $X$ quadrature (e.g.~by using a second cavity coupled to the resonator). The measurement signal from this second measurement, $I_2(t)$, is not subject to squashing because it is not part of the feedback loop.  As a result, the measurement noise in $I_2(t)$ is completely uncorrelated with the feedback signal. Since the second measurement is also a QND measurement of the $X$ quadrature, it does not affect the results for $V_{X,{\rm tot}}^{\mbox{\scriptsize fb}}$ or $S^{\mbox{\scriptsize fb}}_{X,{\rm meas}}(\omega)$ derived above. If the rate of the second measurement is $\tlam$, then the spectrum of its output  (again, referred back to the oscillator) is
\begin{widetext}
\begin{equation} 
	 S^{\mbox{\scriptsize fb}}_{X,{\rm meas,2}} (\omega) \equiv
	 \frac{S^{\mbox{\scriptsize fb}}_{I_2}(\omega)}  {\tlam} =
	\frac{1}{\tlam} + \frac{4}{\gamma}
	\left[
		\frac{(\nbar + 1/2) - \mathcal{A} }
			{ \left( 2 \omega / \gamma \right)^2 + (1+\alpha)^2 }
	+
		\frac{\mathcal{A} }
			{    \left( 2 \omega / \gamma \right)^2 +
				\left(1+ 2 \tkay V_X / \gamma \right) ^2 }
	\right]
	\label{eq:SIsecondmeas}
\end{equation} 
\end{widetext}
where:
\begin{eqnarray}
	\mathcal{A} & = &
		\alpha \frac{ (2 \nbar + 1 )(1 + 2 \tkay V_X / \gamma + \alpha) - \alpha V_X }
		{ \alpha(2 + \alpha) - 2 (2 \nbar+1)\tkay }
\end{eqnarray}
The first term in the spectrum above represents the added noise of the measurement (e.g.~shot noise), while the terms in square brackets are a direct measure of the oscillator's $X$ quadrature fluctuations.  We see that with feedback, these are described by the sum of two Lorentzians.  The integral of the area under these peaks directly yields $V_{X,{\rm tot}}^{\mbox{\scriptsize fb}}$, the total (unconditional) $X$ quadrature variance in the presence of feedback:
\begin{eqnarray}
	 \int \frac{ d \omega}{2 \pi} \left(  
		S^{ \mbox{\scriptsize fb} }_{X,{\rm meas,2} } (\omega) -
	\frac{1}{\tlam} \right) = V_{X,{\rm tot}}^{\mbox{\scriptsize fb}} 
\end{eqnarray}
It thus follows from Eq.(\ref{eq:VXtotfb}) that in the strong feedback limit ($\alpha \ra \infty$), the area under the resonant peak in the output spectrum directly yields $V_X$, and hence a direct
measure of squeezing.
Note that in this limit only the second Lorentzian term in
$S^{\mbox{\scriptsize fb}}_{X,{\rm meas,2}} (\omega)$ survives, as one has:
\begin{eqnarray}
	\lim_{\alpha \ra \infty} \mathcal{A} = (2 \nbar + 1) - V_X
		= V_X \left(1 + 2 \tkay V_X / \gamma \right)
\end{eqnarray}
where we have made use of Eq.~(\ref{eq:VXtot}).  
Thus, for strong feedback, the squeezing of the oscillator can now be unambiguously detected in the output signal of the second measurement:  one obtains a simple Lorentzian resonance whose area is simply $V_X$.  We remind the reader that in the same limit, the spectrum of the first measurement shows no signature of the oscillator.  Note that in practice, the limit of strong feedback is already achieved when $\alpha \gg \max(1, \tkay/ \gamma = 4 \nBA)$. 

\section{Details of Calculations} 
\label{sec:Details}

\subsection{Spectrum of the Detector Output}

\subsubsection{Equations of motion}

The Heisenberg equations of motion (in
the rotating frame) follow directly from $H_{0}$ and the
dissipative terms in the total Hamiltonian:
\begin{subequations}
\begin{eqnarray}
    \dot{ \hat{d} } & = &
        - \frac{\kappa}{2} \hat{d} - \sqrt{\kappa} \hat{\xi}(t) e^{i \omegar t}
        - i \tilde{A} \left[ 
        		c \left(1 + e^{-2 i \omega_M t} \right)
		 +  h.c.
		 \right] \nonumber \\ 
	& = &
        - \frac{\kappa}{2} \hat{d} - \sqrt{\kappa} \hat{\xi}(t) e^{i \omegar t}
        \nonumber \\
        && - i \sqrt{2} \tilde{A} \left[ 
        		\hX \left(1 + \cos(2 \omega_M t) \right) +
			\hY  \sin(2 \omega_M t )
		\right] 
			\label{eq:dEOM} \\
    \dot{ \hat{c} } & = &
        - \frac{\gamma}{2} \hat{c} - \sqrt{\gamma} \hat{\eta}(t) e^{i \omega_M t}
        - i  \tilde{A} \left(1 + e^{2 i \omega_M t} \right)
        \left( \hd + \hd^{\dagger} \right) 
        \nonumber \\
        \label{eq:cEOM}
\end{eqnarray}
\end{subequations}
Here, $\hat{\xi}$ describes noise in the cavity input operator $\bin$.  In the limit where there is only quantum noise (i.e. shot noise) in the cavity drive, we have:
\begin{subequations}
\begin{eqnarray}
	  \langle \hat{\xi}^{\dag}(t) \cdot \hat{\xi}(t') \rangle
	  	& = & 0  \\
	  \langle  \hat{\xi}(t) \cdot  \hat{\xi}^{\dag}(t') \rangle
	  	& = &  \delta(t-t')
\end{eqnarray}
	\label{eq:XiNoise}
\end{subequations}
In contrast, $\hat{\eta}$ describes equilibrium noise due to the intrinsic damping of the mechanical oscillator.  One has:
\begin{subequations}
\begin{eqnarray}
	  \langle \hat{\eta}^{\dag}(t) \cdot \hat{\eta}(t') \rangle
	  	& = & \nbar \delta(t-t') 
		 \\
	  \langle  \hat{\eta}(t) \cdot  \hat{\eta}^{\dag}(t') \rangle
	  	& = & \left( \nbar + 1 \right) \delta(t-t')
\end{eqnarray}
	\label{eq:EtaNoise}
\end{subequations}
where $\nbar$ is a Bose-Einstein occupation factor evaluated at energy  $\hbar \omega_M$ and temperature $T_{\rm bath}$.

The equations of motion are easily solved by first writing them in terms of the quadrature operators $\hX$ and $\hY$, and then Fourier
transforming.  To present these solutions, we first introduce the cavity and mechanical oscillator susceptibilities as:
\begin{subequations}
\begin{eqnarray}
	\chiR(\omega) & = & \frac{1}{-i \omega + \kappa/2} \\
	\chiM(\omega) & = & \frac{1}{-i \omega + \gamma/2 },
\end{eqnarray}
\end{subequations}
and define the back-action force $\hf_{BA}$ via
\begin{eqnarray}
	\hf_{BA}(\omega) = 
		-\tilde{A} \sqrt{2 \kappa} \chiR(\omega) \left(
			\hat{\xi}(\omega+\omega_R) + \hat{\xi}(\omega-\omega_R)
		\right).
\end{eqnarray}
Note that while $\hat{\xi}$ describes white noise, the cavity susceptibility $\chiR(\omega)$ ensures that $\hf_{BA}(\omega)$ is only significant around a narrow bandwidth centered about zero frequency.  Note also that we define Fourier transformed operators via:
\begin{subequations}
\begin{eqnarray}
	\hat{A}(\omega) & \equiv & \int_{-\infty}^{\infty} dt
		\hat{A}(t) e^{i \omega t} \\
	\hat{A}^{\dag} (\omega) & \equiv & \int_{-\infty}^{\infty} dt
		\left[ \hat{A}(t) \right]^{\dag} e^{i \omega t} 
\end{eqnarray}
\end{subequations}
As such, one has $\left[\hat{A}(\omega) \right]^{\dag} = \hat{A}^{\dag}(-\omega)$.

The solutions of the Fourier-transformed quadrature operators then read:
\begin{subequations}
\begin{eqnarray}
    \hX(\omega) & = &
            \chi_M(\omega) 
             \Bigg[
        - \sqrt{ \frac{\gamma}{2 } }
                 \left(  \hat{\eta}(\omega+\omega_M)
             +   \hat{\eta}^{\dag}(\omega-\omega_M) \right)
	\nonumber \\
        && 
		+ \frac{
			\hf_{BA}(\omega+ 2 \omega_M) - 
			\hf_{BA}(\omega- 2 \omega_M)
			}{2 i}
            \Bigg]
            \label{eq:XSoln} \\
    \hY(\omega) & = &
          \chi_M(\omega)  \Bigg[
             i \sqrt{ \frac{\gamma}{2} }  \left(
             \hat{\eta}(\omega+\omega_M)
        -  \hat{\eta}^{\dag}(\omega-\omega_M) \right)
            \nonumber \\
	&&
	-\hf_{BA}(\omega) - \frac{
		 \hf_{BA}(\omega + 2 \omega_M) + \hf_{BA}(\omega- 2 \omega_M)
		 	}{2}
           \Bigg]    \nonumber 
         \\  \label{eq:YSoln}
\end{eqnarray}
	\label{eq:XYSols}
\end{subequations}
Note from Eqs.~(\ref{eq:XYSols}) that there is no back-action damping of either quadrature, even when one deviates from the good cavity limit by having $\kappa / \omega_M > 0$.  This is easy to understand on a purely classical level.  Note first that that it is only the cosine quadrature (i.e.  $\hd + \hd^{\dag}$) of the cavity which couples to the mechanical resonator.  As the cavity is itself a harmonic oscillator, this means that only the cavity sine quadrature (i.e. $\hd - \hd^{\dag}$) will be affected by the resonator motion.  As the cavity cosine quadrature provides the back-action force on the resonator
(c.f. Eq.~(\ref{eq:CavityForceOp})), it thus follows that the back-action force is {\em completely} independent of both quadratures of the mechanical resonator's motion.  There is thus no back-action damping, as such damping requires a back-action force which responds (with some time-lag) to the motion of the oscillator.  

Equations (\ref{eq:NoiseSpectra}) for the noise spectra of $\hX$ and $\hY$ at frequencies $\omega \ll \kappa$ now follow directly from Eqs.~(\ref{eq:XYSols}) and Eqs.~(\ref{eq:XiNoise}),(\ref{eq:EtaNoise}) which determine the noises $\hat{\xi}$ and $\hat{\eta}$.

\subsubsection{Output Spectrum and Beating the SQL}

Standard input-output theory \cite{WallsMilburn, Gardiner00} yields the following relation between $\hat{b}_{\rm out}$, the field leaving the cavity, and $\bin$, the field entering the cavity:
\begin{eqnarray}
	\hat{b}_{\rm out}(t) = \hat{b}_{\rm in}(t) + \sqrt{\kappa} \hat{a}(t)
\end{eqnarray}
In our case of a one-sided cavity, this relation becomes in the lab (i.e.~non-rotating) frame:
\begin{eqnarray}
    \bout(\omega) & = &
        \bar{b}_{\rm out}(\omega)+
        \left[
            \frac{ -i (\omega-\omegar) - \kappa/2}
            { -i (\omega-\omegar) + \kappa/2} \right] \hat{\xi}(\omega)
    \nonumber \\
    &&
        - i \tilde{A} \sqrt{2 \kappa} \chiR(\omega-\omegar) \cdot
        \hX(\omega-\omegar)
                     \label{eq:DSBbout}
\end{eqnarray}
The first term on the RHS simply represents the output field from
the cavity in the absence of the mechanical oscillator and any
fluctuations.  It will yield sharp peaks at the two sidebands
associated with the drive, $\omega = \omegar \pm \omega_M$.
The second term on the RHS of Eq.~(\ref{eq:DSBbout}) represents
the reflected noise of the incident cavity drive.  This noise will
play the role of the ``intrinsic output noise" or ``measurement imprecision" of this detector.

Finally, the last term on the RHS of Eq.~(\ref{eq:DSBbout}) is the
amplified signal: it is simply the amplified quadrature
$X$ of the oscillator.  We see that the dynamics of $\hX$ will result in a signal of bandwidth $\sim \gamma$ centered at the cavity resonance frequency.  This can be detected by making a homodyne measurement of the signal leaving the cavity.  Using a local-oscillator amplitude $b_{LO}(t) = i B e^{-i \omega_R t}$ with $B$ real, and defining the homodyne current as:
\begin{eqnarray}
	\hat{I}(t) = \left( b_{LO}^*(t) + \hat{b}_{out}^{\dag}(t) \right)
		\left( b_{LO}(t) + \hat{b}_{out}^{\pd}(t) \right)
\end{eqnarray}
one finds that the fluctuating part of $I$ is given in frequency-space by:
\begin{eqnarray}
	\hat{I}(\omega) & = & -B \Big[
		2 \sqrt{2} \tilde{A} \sqrt{ \kappa} \chiR(\omega)
		 \hX(\omega)   +
		\label{eq:IHomo}
		 \\
		&&
		i \frac{i \omega + \kappa/2}{i \omega - \kappa/2}							 		\left(
				\hat{\xi}(\omegar+\omega) - \hat{\xi}^{\dag}(-\omegar+\omega) 
			\right)
			\Big]
	\nonumber
\end{eqnarray}
The signal associated with the oscillator will be in a bandwidth $\sim \gamma \ll \kappa$: for these frequencies, the above expression simplifies to:
\begin{eqnarray}
	\hat{I}(\omega) & = & -B \Big[
		\frac{4\tilde{A}} { \sqrt{\kappa/2} } 
		 \hX(\omega)   +
		\label{eq:IHomoSimp}
		 \\
		&&
		-i	
			\left(
				\hat{\xi}(\omegar+\omega) - \hat{\xi}^{\dag}(-\omegar+\omega) 
			\right)
			\Big] \nonumber
\end{eqnarray}

Using this equation along with Eqs.~(\ref{eq:XSoln}), (\ref{eq:XiNoise}) and (\ref{eq:EtaNoise}), it is straightforward to obtain the result for the homodyne spectrum $S_{I}(\omega)$ given in Eq.~(\ref{eq:HomoSpectrum}).

\subsection{Conditional Evolution}

To derive the stochastic master equation describing the conditional evolution of the resonator under the double sideband measurement scheme, (that is, the evolution given the continuous stream of information obtained by the observer), one uses a procedure that is essentially identical to that given in Ref.~\cite{DJ}. Once we have moved into the interaction picture (in which the quadratures are QND observables), the displacement picture~\cite{WM93} (that is, separated $\hat{a}$ into $\bar{a}$ and $\hat{d}$ as per Eq.~(\ref{eq:DisplacementTrans})), and made the rotating-wave approximation, the Hamiltonian for the combined cavity and resonator system is 
\begin{equation}
    H = -\sqrt{2} \tA (\hd + \hd^\dagger) X .
\end{equation}
We now perform homodyne detection of output from the (one-sided) cavity, and as a result the evolution of the system is given by the quantum optical stochastic master equation~\cite{Carm93,WM93}
\begin{equation}
 d\sigma = -\frac{i}{\hbar}
[H,\sigma] dt + \kappa{\cal D}[\hd]
\sigma dt +
\sqrt{\eta\kappa} {\cal H}[-i\hd] \sigma  dW,
  \label{homsme}
\end{equation}
where $\sigma$ is the joint density matrix of the two systems as before $\eta$ is the detection efficiency, and $\kappa$ is the cavity decay rate. The superoperators ${\cal D}$ and ${\cal H}$ are given by 
\begin{eqnarray}
2{\cal D}[\hc] \sigma & = & 2 \hc \sigma_{\mbox{\scriptsize
c}} \hc^\dagger - \hc^\dagger \hc \sigma -
\sigma \hc^\dagger \hc , \\ {\cal H}[\hc]
\sigma & = & \hc \sigma +
\sigma \hc^\dagger - \mbox{Tr}[\hc
\sigma + \sigma \hc^\dagger ]
\sigma ,
\end{eqnarray}
for an arbitrary operator $\hc$. 

We now wish to obtain an equation for the evolution of the resonator alone. This is possible so long as the cavity decay rate is fast compared to the timescale of the cavity-resonator interaction. That is, 
\begin{equation}
   \frac{\tA \sqrt{\langle X^2 \rangle}}{\kappa} \sim \frac{\gamma}{\kappa}
  \equiv \epsilon \ll 1 ,
\end{equation}
This means that the light ouput from the cavity spends sufficiently little time in the cavity that it continually provides up-to-the-minute information about the oscillator. With this large damping rate, the fluctuations of the light in the cavity about the average value $\bar{a}$ are small, and we can thus expand the cavity state described by the operator $\hd$ about the vacuum:
\begin{eqnarray}
\sigma & = &
\;\;\; \rho_{\mbox{\scriptsize 00}}
|0\rangle\langle 0| + (\rho_{\mbox{\scriptsize
10}}  |1\rangle\langle 0| + \mbox{H.c.}) \nonumber \\
& & + \rho^{\mbox{\scriptsize
a}}_{\mbox{\scriptsize 11}}  |1\rangle\langle 1| +
(\rho_{\mbox{\scriptsize 20}}  |2\rangle\langle 0|
+ \mbox{H.c.}) + O(\epsilon^3) .
\end{eqnarray}
The density matrix for the resonator is then given by 
\begin{equation}
  \rho = \mbox{Tr}_{\mbox{\scriptsize c}} [\sigma] = \rho_{\mbox{\scriptsize 00}}  
                                                   + \rho_{\mbox{\scriptsize 11}}  + O(\epsilon^3).
\end{equation}
where $\mbox{Tr}_{\mbox{\scriptsize c}}$ denotes the trace over the 
cavity mode. 
From the master equation (Eq.(\ref{homsme})) we then derive the equations of motion for the $\rho_{ij}$. Adiabatic elmination of the off-diagonal elements $\rho_{01}$ and $\rho_{02}$ (described in detail in Ref.\cite{DJ}) allows us to write a closed set of equations for the diagonal elements $\rho_{00}$ and $\rho_{11}$. The result is a  stochastic master equation for $\rho = \rho{00} + \rho{11}$, which is  
\begin{eqnarray}
  d\rho & = & k [X, [X,\rho]] dt + \sqrt{2 \eta k}{\cal H}[X] \rho dW   ,
\label{smepm}
\end{eqnarray}
where the measurement strength $k = 4 \tA^2 / \kappa$. Defining $\tkay = 8 \eta k$, and making a Gaussian ansatz for the quantum state, we find Eqs.~(\ref{eq:est1}) - (\ref{eq:est5}) for the means and variances of the quadratures $X$ and $Y$.

\subsection{Squeezing via Feedback Control}  

There are three formulations that can be used to analyze the behavior of an observed linear quantum system: the Heisenberg picture (the input-ouput formalism), the Schr\"{o}dinger picture (the SME) and the equivalent classical formulation, introduced in Ref.~\cite{DJ}. We have already used the first two methods in our analysis above. To analyze the effect of  feedback we now use the third. The equivalent classical formulation is given by the equations 
\begin{eqnarray}
    dx & = & -\frac{\gamma}{2} x dt+ \sqrt{\gamma\tTeq} dW_x    \\
    dy & = & -\frac{\gamma}{2} y dt + \sqrt{\gamma\tTeq} dW_y + \frac{\sqrt{\tkay}}{2} dV_1  +  \frac{\sqrt{\tlam}}{2} dV_2  
\end{eqnarray}
where $x$ and $y$ are now classical dynamical variables, and as always the noise sources, $W_i$ and $V_i$, are mutual uncorrelated Wiener processes. We have now included two measurements of the $x$ quadrature, one with strength $\tkay$ and the other with strength $\tlam$, for reasons that will be explained below. The measurement records (i.e.~the homodyned output signals) for these measurements are given by 
\begin{eqnarray}
 dI_1  & = &  \sqrt{\tkay} x dt + dU_1  \\ 
 dI_2  & = &  \sqrt{\tlam} x dt + dU_2  .
\end{eqnarray}
Once again the $U_i$ are mutually uncorrelated Wiener processes. Of interest are the quantities $\langle \bar{X}\rangle_1$ and  $\langle \bar{X}\rangle_2$, which are (respectively) the two  observers' estimates of the $X$ quadrature.  Note that these are {\it not} the same as $x$ above.  When $\tlam=0$, so that there is no second measurement, the equation of motion for $\langle \bar{X}\rangle_1$ is naturally  that given by Eqs (\ref{eq:est1}) - (\ref{eq:est5}). With the second measurement, the dynamics of the means and variances for the first observer become 
\begin{subequations}
\begin{eqnarray}
   \!\!\!\!\!\!\!\!     d\barx_1  & = & -(\gamma/2) \barx dt + \sqrt{\tkay} V_X  d\tilde{U}_1 \\
   \!\!\!\!\!\!\!\!	d\bary_1 & = & -(\gamma/2) \bary dt + \sqrt{\tkay} C  d\tilde{U}_1 \\
   \!\!\!\!\!\!\!\!      \dot{V}_X & = & - \tkay V_X^2 - \gamma (V_X - \tTeq)  \\
   \!\!\!\!\!\!\!\!	\dot{V}_Y & = & - \tkay C^2  + 2 k + 2\lambda  
                                                   - \gamma( V_Y - \tTeq) \\  
   \!\!\!\!\!\!\!\!	\dot{C}\; & = & - \gamma C  - \tkay V_X C   \label{eq:est5b}
\end{eqnarray}
\end{subequations}
where $k = \tkay/(8\eta_1)$ and $\lambda = \tlam/(8\eta_2)$ are the strengths of the respective measurements (under the usual definition of measurement strength~\cite{JacobsSteck06}), and the $\eta_i$ are the respective efficiencies of the measurements. We also introduce a fourth set of noises, $\tilde{U}_i$, where $\tilde{U}_1$ appears in the above equations for the first observer, and $\tilde{U}_2$ would appear in the equations for the second observer, although we will not need those here. The $\tilde{U}_i$ are given by~\cite{DJ} 
\begin{eqnarray} 
 	d\tilde{U}_i  & = &  dI_i - \barx_i dt =  \sqrt{\tkay} (x -\barx_i ) dt + dU_i  . 
 \label{eq:u2u}
\end{eqnarray} 
While it is not obvious, it turns out that the $d\tilde{U}_i$ are also mutually uncorrelated, and uncorrelated with all the other noise sources.

Armed with the above equations, we now introduce feedback into the system. We apply a continuous feedback force $F(t) = \alpha \gamma \bar{X} \sin(\omega_M t) $ to the system in the lab frame. Discarding  rapidly oscillating terms (making a rotating-wave approximation), this results in the following dynamics for the system: 
\begin{eqnarray} 
    dx & = & -\left( \frac{\gamma}{2} x  +  \frac{\alpha \gamma}{2} \barx_1 \right) dt + 
    	\sqrt{\gamma\tTeq} dW_x    \\
    dy & = & - \frac{\gamma}{2} y dt 
    + \sqrt{\gamma\tTeq} dW_y 
               \nonumber \\ & & + \sqrt{\tkay/4} dV_1  +  \sqrt{\tlam/4} dV_2  
\end{eqnarray} 
The feedback provides a damping force on the $X$ quadrature with a rate $\alpha \gamma/2$. Since applying a known force to the system cannot change the observers' uncertainty regarding the classical coordinates, the equations of motion for the variances for both observers remain the same. The equations of motion for the means however also pick up exactly the same damping terms. Thus for observer one we have 
\begin{eqnarray} 
        \dot{ \barx}_1  & = & -  \frac{(1+\alpha) \gamma}{2} \barx_1 + \sqrt{\tkay}
        	 V_X  \dot{ \tilde{U} }_1 \\
	\dot{\bary}_1 & = & - \frac{ \gamma}{2}\bary_1 + \sqrt{\tkay} C  \dot{ \tilde{U} }_1 .
\end{eqnarray} 

We now want to calculate the variance of the $X$ quadrature under this feedback protocol, and also the spectrum of the output signal for both observers. Since the $X$ and $Y$ quadratures are not coupled, we need merely solve the two coupled equations 
\begin{eqnarray} 
 \!\!\!\!\!\!\!    \dot{x}  \!\! & = & \!\!  -\frac{\gamma}{2} x  - \frac{\alpha \gamma}{2} \barx_1 + \sqrt{\gamma\tTeq} \dot{W}_x   \label{eq:dx} \\
  \!\!\!\!\!\!\!    \dot{\barx}_1 \!\! & = & \!\!  - \! \left( \frac{\gamma}{2} + \frac{\alpha \gamma}{2} + \tkay V_X \right) \! \barx_1 + \tkay V_X x  + \sqrt{\tkay} V_X  \dot{U}_1 ,
     \label{eq:dX1}
\end{eqnarray} 
where we have used Eq.(\ref{eq:u2u}) to write the equation for $\barx_1$ in terms of $\dot{U}_1$ rather than $\dot{ \tilde{U} }_1$. The unconditional variance of the $X$ quadrature under feedback, which we will denote by $V_{X,{\rm tot}}^{\mbox{\scriptsize fb}}$, is given by the variance of $x$.  Solving for the steady-state value of $V_{X,{\rm tot}}^{\mbox{\scriptsize fb}}$ using the usual techniques of Ito calculus, and using the fact that $\tTeq = V_X + (\tkay/\gamma)V_X^2$ (c.f. Eq.~(\ref{eq:VXtot})), we obtain Eq.~(\ref{eq:VXtotfb}).
We see that as the feedback strength $\alpha$ tends to infinity, $V_X^{\mbox{\scriptsize fb}} \rightarrow V_X$, as claimed above. 

To calculate the spectrum of the output signal for the first observer we first transform Eqs (\ref{eq:dx}) and (\ref{eq:dX1}) to the frequency domain and solve them. The solution is of the form
\begin{eqnarray}
     \left( \begin{array}{c} x(\omega) \\ \barx_1(\omega) \end{array} \right) = M(\omega) \left( \begin{array}{c} \sqrt{\gamma\tTeq} \dot{W}_x (\omega) \\ \sqrt{\tkay} V_X  \dot{U}_1 (\omega)  \end{array} \right) 
	\label{eq:Meqn}
\end{eqnarray}
with 
\begin{equation} 
    \langle \dot{W}_x (\omega) \dot{W}_x (\omega')\rangle = 
    \langle  \dot{U}_1 (\omega)  \dot{U}_1 (\omega') \rangle = \delta(\omega + \omega') .
	\label{eq:NoiseCorrelators}
\end{equation}
The output signal for the first measurement is:
\begin{equation}
   I_1(\omega) = \sqrt{\tkay} x(\omega) + \dot{U}_1 (\omega) . 
\end{equation}
The corresponding output spectrum is defined via:
\begin{equation}
	\langle I_1(\omega) I_1(\omega') \rangle = 
	S_{I_1}(\omega) \delta(\omega + \omega')
\end{equation}
Using Eqs.~(\ref{eq:Meqn}),(\ref{eq:NoiseCorrelators}), one finds the zero-frequency spectrum to be given by:
\begin{equation}
   S_{I_1}(0) =  \frac{(1+\alpha+ 2 \tkay V_X / \gamma)^2}{(1+\alpha)^2}   . 
\end{equation}
where we have made use of Eq.~(\ref{eq:VXtot}).  Referring this back to the oscillator by dividing by $\tkay$ results in Eq.~(\ref{eq:SI1fb}).

Similarly, the output of the second measurement is given by 
\begin{equation}
   I_2(\omega) = \sqrt{\tlam} x(\omega) + \dot{U}_2 (\omega) ,  
\end{equation}
Using this definition and Eqs.~(\ref{eq:Meqn}),(\ref{eq:NoiseCorrelators}), and making use of 
Eq.~(\ref{eq:VXtot}), we find the output spectrum given in Eq.~(\ref{eq:SIsecondmeas}).

\section{Conclusions}
\label{sec:Conc}

In this paper, we have provided a thorough and fully quantum treatment of back-action evasion using a driven electromagnetic cavity which is parametrically coupled to a mechanical oscillator.  We have considered both the unconditional and conditional aspects of the measurement.  In particular, we have derived exactly how strong the coupling must be to beat the standard quantum limit, and to achieve a conditionally squeezed state.  We have also shown how feedback can be used to generate true squeezing, and how this squeezing can be detected using a second measurement.

\section*{Acknowledgments}

We thank S. Girvin and K. Schwab for useful conversations.  A.C. acknowledges the support of NSERC, the Canadian Institute for Advanced Research, and the Alfred P. Sloan Foundation.  F.M. acknowledges support by NIM, SFB 631, and the Emmy-Noether program of the DFG.

\bibliographystyle{apsrev}
\bibliography{ACrefs}

\end{document}